# The Planck Telescope

F.Villa[1], M.Bersanelli[2], C.Burigana[1], R.C.Butler[1], N.Mandolesi[1], A.Mennella[3], G.Morgante[1], M.Sandri[1], L.Terenzi[1], L.Valenziano[1]

[1] *Istituto TESRE/CNR – Bologna – Italy*
[2] *Università di Milano – Milano – Italy*
[3] *Istituto di Fisica Cosmica / CNR – Italy*
*On behalf of the Planck Collaboration*

**Abstract.** In this paper we present an overview of the Telescope designed for ESA's mission dedicated to map the Cosmic Microwave Background Anisotropies and Polarization. Two instrument, LFI and HFI, operate in an overall frequency range between 25 and 900 GHz and share the focal region of the 1.5 meter optimized aplanatic telescope. The optimization techniques adopted for the optical design and the telescope characteristic are reported and discussed.

## INTRODUCTION

Planck is a mission dedicated to imaging the anisotropies of the Cosmic Background Radiation with a typical angular resolution of 10 arcmin and an average sensitivity per pixel of 2 – 4 ?K/K. Two instrument, the radiometric Low Frequency Instrument (LFI) [1] and the bolometric High Frequency Instrument (HFI) [2], are coupled to an optimized dual reflector off–axis telescope with a 1.5 meter of projected aperture. Both instrument, cryogenically cooled respectively at 20 K and 100 mK, will measure the radiation coming from the sky and scattered by the telescope between 30 GHz and 100 GHz for LFI and between 100 GHz and 857 GHz for HFI. The telescope is passively cooled at about 50 K by a thermal design of the payload based on a set of three dedicated V–grooved shields. The electromagnetic design of the telescope, carried out after approximately three years of intensive studies and simulations, has been consolidated during the Planck Payload Architect industrial activity, kicked off on Jan 99 and completed on Dec 99. The study was performed by ALCATEL, under the control of ESA, with a strong scientific and technical support of the instrument teams, and of the Telescope Provider.

An accurate study of the telescope performances plays a fundamental role in the understanding of the systematic effects in Planck. Any non ideality in the telescope performance will contaminate the measurements to some extent. As a consequence the arising effects must be analyzed and understood in detail. The non–ideal behavior can be divided in two main categories: the aberration of the main beam which degrade the angular resolution [3] [4] and the near and far side lobes which contribute to the Straylight Induced Noise, or briefly SIN [5]. In this work a description of the Planck telescope is given, emphasizing the evolution of the design from the first layout to the present baseline. The electromagnetic performances are briefly discussed.

## THE PLANCK TELESCOPE

The Telescope is based on a two–mirror off–axis scheme (see Figure 1) which offers the advantage to accommodate large focal plane instruments with an unblocked aperture an thus maintaining the diffraction by the secondary mirror and struts at very low levels. The field of view is as large as +/– 5° and centered on the line of sight (LOS) which is tilted at about 3.7 degrees with respect to the main reflector axis. The LOS is 85° away from the spin axis which points in the anti – sun direction.

The Telescope will work in the frequency range between 25 GHz to 1 THz. In this range the total emissivity of the telescope will be less than 1 % at the beginning and less than 5% at the end of the mission.

Both the primary and the secondary mirrors have an ellipsoidal shape as in the case of the Gregorian aplanatic design. The sub reflector revolution axis is tilted with respect to the main reflector revolution axis. The primary mirror physical dimensions are about 1.9 x 1.5 meters, allowing a projected circular aperture of 15 meter of diameter. The secondary reflector has been oversized up to approximately 1 meter of diameter to avoid any additional under illumination of the primary. LFI and HFI are located in the focal region which is approximately a 8° tilted plane with respect to the plane perpendicular to the $Z_{RDP}$. The typical roughness of the mirror surface is required to be 1 ?m at any scale up to 0.8 mm and the average mechanical surface error will be 10?m with respect to the best fit surfaces (2?m of amplitude maximum for periodic structures).

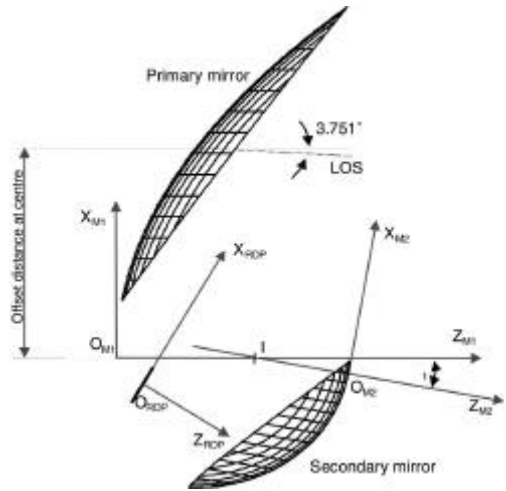

**FIGURE 1.** Layout of the Planck Telescope. The Line of Sight is tilted at 3.751 degree with respect to the primary mirror axis. The secondary mirror axis is tilted at 10.1 degrees with respect to the primary mirror axis. The main coordinate systems are indicated: the primary reflector frame (M1), the secondary reflector frame (M2) and the Reference Detector Plane (RDP) which represents the center of the focal plane. The Line of Sight (LOS) is also showed. $Z_{M1}$ and $Z_{M2}$ are the revolution axes of the primary and secondary ellipsoids, respectively.

The Planck Telescope as a complete satellite sub–unit (see Figure 2) includes the primary reflector, the primary mirror supporting structure, the secondary reflector, the secondary reflector fixation struts, the telescope frame (holding the two mirror support structures, the focal plane array and the interfaces with the payload module struts), the payload module straylight baffle (interfaced with the coldest V–groove shield at 50 K), the telescope inner baffle between the focal plane array and the secondary mirror, the primary reflector extension baffle, the instrumentation for the telescope hardware, and the hardware to interface with the instruments.

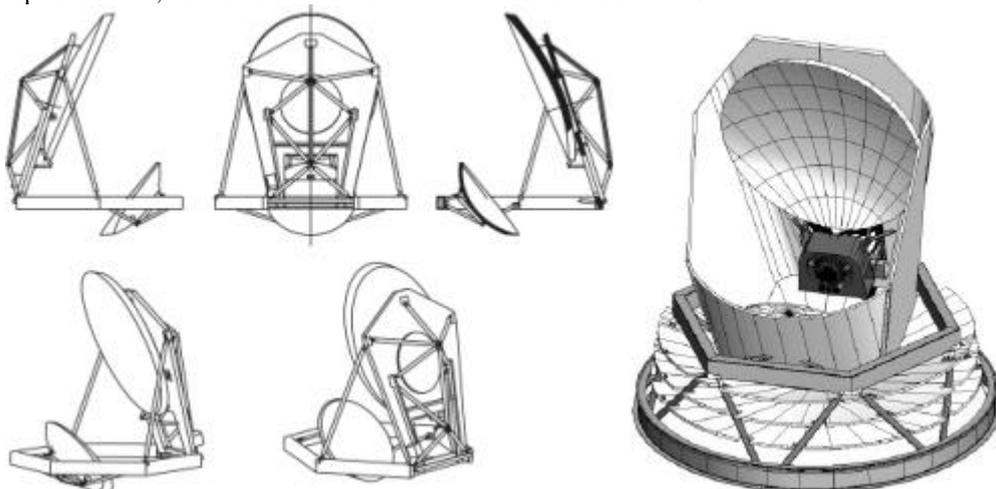

**FIGURE 2.** Left: the Planck Telescope as a single unit to be integrated in the payload module. Right: the Planck payload and the telescope. The V-grooves are seen as well as the focal plane array accommodating the two instruments. The hexagonal structure, visible on the top of the third V-grooved shields, is the telescope frame.

The primary and secondary mirrors will be fabricated using Carbon Fiber (CFRP) technology. The baseline consists of an all-CFRP rounded triangular tubes sandwich array arranged in a honeycomb–like structure. This kind of structure has been chosen to satisfy the requirements of low mass (< 120 Kg including struts and supports), high stiffness, high dimensional accuracy, and low thermal expansion coefficient. The sandwich concept consists of a thick (4-10 cm) honeycomb-like core with the desired shape, and to which are bonded two thin (1-1.5 mm) reflecting skins.

## The History

The current Planck Telescope Configuration is the result of an evolution of several designs started from the one reported in the COBRAS proposal in 1993 [6]. The COBRAS telescope was a "*clear – aperture, 0.6 m focal length Gregorian with primary parabolic reflector of 1.5 m diameter (1.0 meter illuminated) and an elliptical secondary of 0.57 m diameter*". The design was modified at the time of the Phase–A proposal in 1996 [7] when COBRAS and SAMBA were joined in a single mission. A 1.3 meter projected aperture off-axis telescope satisfying the Dragone–Mizuguchi (DM) condition [8] was chosen for COBRAS/SAMBA. The DM configuration exhibits an ideal response only in the center of the focal region. In this point the telescope is equivalent to a blockage free on–axis configuration. Unfortunately, as the LFI optical working group demonstrated, the aberrations (especially the coma) increase significantly for the feeds located outside the center of the focal surface [9]. The degradations of the telescope performances (mainly the effective angular resolution) in the main beam region affected the scientific capabilities of LFI, whose feeds are located in a ring around HFI with a typical beam location approximately 3 degrees away from the telescope optical axis. After the announcement of opportunity for the FIRST/Planck program, in February 1998, the telescope design was increased in projected aperture from 1.3 to 1.5 meters of diameter. To avoid any change of the focal plane design only the main reflector was changed improving the stray light rejection by adding a ring around the top edge of the primary mirror. The so called "*Carrier Configuration*" was in fact a 1.3 – like telescope with a more under illuminated 1.5 meter primary mirror. Several optical configurations were studied and designed by the LFI team and DSRI with the purpose of investigating the best 1.5 meter telescope configuration for Planck.

An aplanatic design was obtained starting from a Gregorian standard configuration by modifying the conical constants and curvature radius according to the analytical "*aplanatic*" condition [10]. The aplanatic choice of the mirror shapes (both ellipsoidal) minimizes the spherical and comatic aberrations even in the case of the off–axis scheme, as readily seen in the left panels of Figure 3 and extensively reported in [11]. Several aplanatic designs were considered and proposed as alternative solutions for the telescope. The improvements on the main beam shape due to the aplanatic configuration are evident also for the off–axis feeds. During the Planck Payload Architect industrial activity on 1999, a telescope optimization process was carried out.

## The Optimization

The optimization [12] has been addressed with the aim to minimize the beam distortions and the straylight noise. The spillover energy and the ellipticity of the main beams have been assumed as the quality evaluation parameters. Moreover, one of the additional parameter used for selecting the best configuration was the flatness of the focal surface in order to minimize the obscuration between the feeds on the focal plane unit. Operatively the study was mainly devoted to the optimization of the mirrors shapes by the minimization of the Wave Front Error (WFE), using the optical software $^{®}$CODEV. Although this software has been conceived for systems in the optical wavelength range, a minimization of the WFE or alternatively the maximization of the Strehl Intensity ratio corresponds to the minimization of the aberrations in a more general context. The advantage is that in $^{®}$CODEV (or similar software) an automatic optimization of the WFE can be possible. However a check of the optimization results was required by the LFI team and a set of simulations of the telescope performances was run using Physical Optics based software like GRASP8. The following parameters have been tuned to optimize the telescope starting from an aplanatic design: the conic constant of both mirrors, the curvature of the sub reflector, the distance and the tilting between the two mirrors, and the FOV direction. The optimization has been performed minimizing the WFE for a total of sixteen (eight for LFI and eight for HFI) different focal points equally distributed in space as reported in the right panel of Figure 3. The so–called CASE1 configuration has been selected as baseline among several optimized designs and it has been shown to be the best compromise between the optical performances and the available room on the payload module.

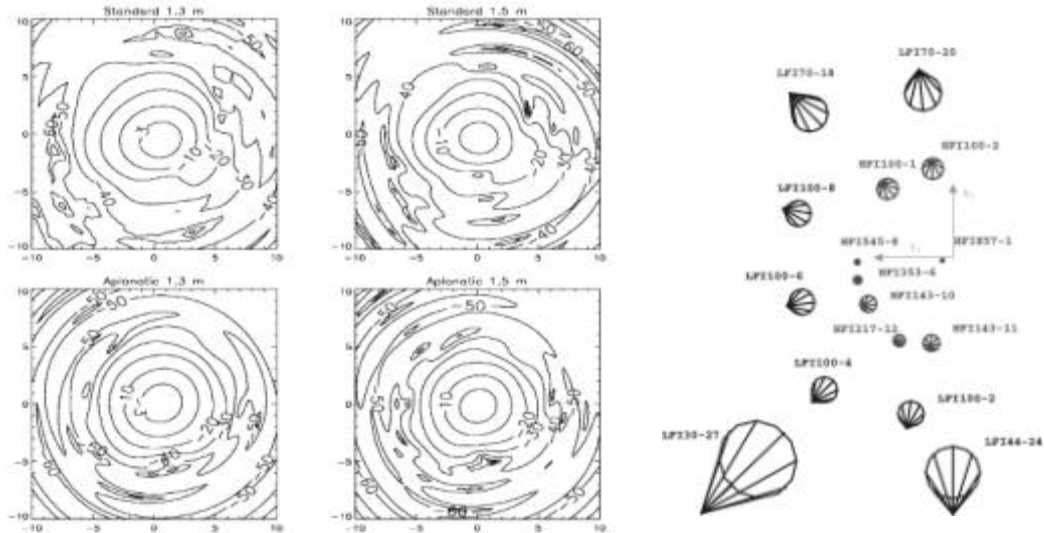

**FIGURE 3.** The four contour plots on the left show the improvement of the beam symmetry due to the aplanatic design proposed by the LFI team. The first row of contour plots reports the main beam response of two Dragone–Mizuguchi different configurations with 1.3 m and 1.5 m of diameter from left to right respectively (Standard 1.3 m and Standard 1.5 m). The second row reports from left to right the response of the aplanatic configurations with 1.3 m and 1.5 m of aperture respectively. These two last configurations were proposed by the LFI team as alternative solutions before starting the optimization of the telescope. All the calculations have been done at TESRE and at 100 GHz with a typical LFI off–axis feed. The right panel shows the feed locations have been used for the optimization.

## The Performances

The CASE1 configuration shows a significant improvement of the telescope performances for all the LFI and HFI detectors concerning the beam shape and the far side lobes, with respect to the Carrier configuration. The optical quality of the optimized telescope has been carefully investigated by the LFI team because of the highly off–axis location (3 degrees on average) of the LFI horns, particularly for the most important cosmological frequency channel at 100 GHz. The left plot of Figure 4 shows the main beam shape on the sky of all of the LFI detectors. It is readily seen that the coma and spherical aberrations are kept at very low levels. The simulated beams show elliptical shape up to the 20 dB contour approximately, which means a really good improvement with respect to the carrier configuration (see the contour plots of Figure 3 and the right panel of Figure 4) even for the most offset beams. As an example of the response of the telescope at the HFI frequencies, the simulated beam shape of one of the 217 GHz HFI channel is reported in the center plot of Figure 4.

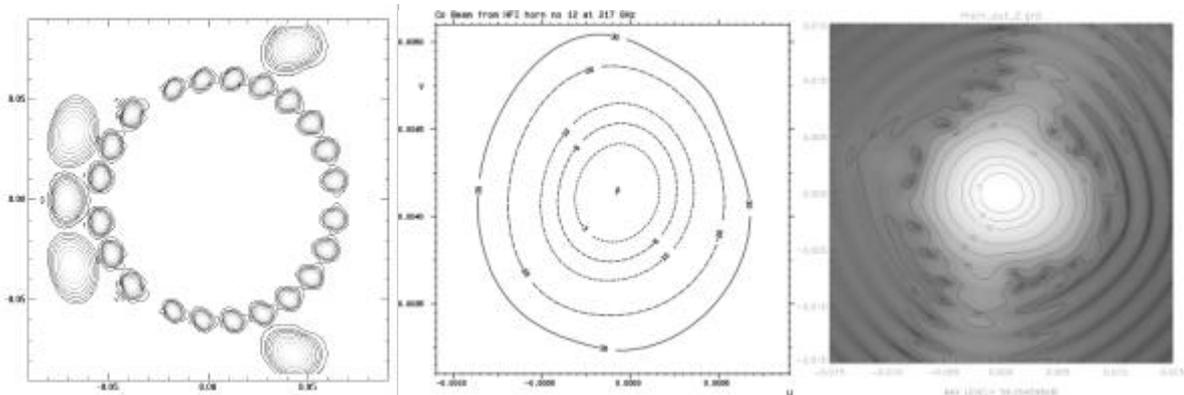

**FIGURE 4.** The performances of the Planck Telescope are summarized in these plots. The contour plots are calculated on the (U,V) plane. Left: the finger print on the sky of the LFI detectors. Center: one of the HFI 217 GHz detector response on the sky. Right: One of the LFI 100 GHz beam. Note that obviously the scale is different for each plot and that the beam symmetry is acceptable even for the most off–axis beam position.

The beam calculation is only the first step for the complete modelization of the telescope which must also include the analysis of roughness, tolerances, misalignments, periodic structure effects, molecular and particles contamination effects, cooldown and thermo–mechanical effects. All the simulations will be supported and verified with an appropriate test campaign on a RF–model of the telescope during 2002.

## CONCLUSIONS

The Planck telescope represents a challenge for telescope technology and optical design. The wide frequency coverage (from 25 GHz to 1000 GHz), the high performances required by both HFI and LFI instruments sharing the 400 x 400 mm wide focal region, and the cryogenic environment (40 – 65 K) in which the telescope will operate, have never been obtained before in experimental cosmology. A comparison between the optical simulations and the measurements at several frequencies and feed locations is mandatory to validate the electromagnetic models used for predicting the performances.

The telescope represents the interface between the sky and the focal plane detectors and a precise study and characterization of its performances is a powerful tool to understand the related systematic effects and to reach the scientific accuracy required in the analysis of the Planck data.

## ACKNOWLEDGMENTS

We wish to thank the European Space Agency (ESA), TICRA engineering consultant, the HFI consortium, the LFI consortium, and the TP consortium.